\documentclass[10pt,a4paper]{article}
\usepackage{amsmath}
\usepackage{amssymb}
\usepackage{amsfonts}
\usepackage{amstext}
\usepackage{amsbsy}
\usepackage[mathscr]{eucal}
\usepackage{graphicx}
\usepackage{color}
\newcommand{\beq}{\begin{equation}}
\newcommand{\eeq}{\end{equation}}
\newcommand{\beqa}{\begin{eqnarray}}
\newcommand{\eeqa}{\end{eqnarray}}
\newcommand{\ket}[1]{| #1 \rangle}

\title{\Large\textbf{Complex Projective Scheme Approach to The Geometrical Structures of Multipartite Quantum Systems}}

\author{\textit{ Hoshang Heydari}\\
        \small\textit{Physics Department, Stockholm university 10691 Stockholm Sweden}\\
\\\small\textit{Email: hoshang@fysik.su.se}}
\begin{document}
\maketitle

\begin{abstract}
In this paper, I will discuss the geometrical structures of multipartite quantum systems based on complex projective schemes. In particular, I will explicitly construct multi-qubit states in terms of these schemes and also discuss separability and entanglement of bipartite and multipartite quantum states. These results give
 some geometrical insight to the fascinating quantum mechanical phenomena of entanglement which is
 fundamentally important and has many applications in the field of quantum computing.
\end{abstract}


\maketitle

\section{Introduction}
Multipartite quantum systems are very interesting  complex composite
systems which are defined on the projective Hilbert spaces.
 The
geometrical structures of
 these complex projective spaces which are directly related to
 separability
and entanglement of quantum states are very important in the field
of quantum information and quantum computing.

   In differential and topological geometry, manifolds are made by
gluing together open balls from Euclidean space.
    In analogy, schemes
are made by gluing together simple open sets which are called affine
schemes.
   However, there is a important difference between these two
constructions.
    In manifold one point looks locally just like
another.
  But schemes admit much more local variation e.g., by
allowing the smallest open sets to be so large that many interesting
geometry can happens within each one of them.

   Affine scheme is an
object that is constructed from commutative ring.
      This is a
generalization  of the relationship between an affine variety and
its coordinate ring.
     For example in the classical algebraic geometry
we have correspondence between the set of affine variety and the
finitely generated nilpotent-free rings over an algebraically closed
field such as complex number field $K=\textbf{C}$.
   In
scheme-theoretic approach there is a correspondence  between the set
of affine scheme and the set of  commutative rings with identity.
    Thus the ring and the corresponding affine scheme are equivalent
objects.
 To give a general definition of scheme we need two abstract objects:
 (1) prime spectrum of a commutative ring Spec $R$, that is
$
  Spec R=\{p:~ p~\emph{is a prime ideal of}~R\}
$
 and (2) sheaf $O_{X}$ where $X$ is a topological space.
  Sheaves are standard objects in many branches of mathematics such as
algebraic topology and geometry.
                The concept of sheaf gives a
systematic way of keeping track of local data on topological space. We assume that readers are familiar with basic of commutative ring and sheaf theory. For those how want to known more about scheme we recommend
  following books \cite{Ueno,Eisen,Mumf,Hart77} which are also our main references.

\section{Quantum systems and schemes}

First we will discuss the geometrical structure of a general single
quantum system $ \ket{\Psi}=\sum^{N_{1}}_{i_{1}=0} \alpha_{i_{1}}
\ket{i_{1}}=\sum^{N}_{i=0} \alpha_{i} \ket{i} $. For an $(N+1)$-dimensional vector space $E$ over a
complex number field $\mathbf{C}$, the spectrum
$\mathbf{P}(E)(\mathbf{C})$ of $\mathbf{C}$-valued points of
$\mathbf{P}(E)$ over $\mathbf{C}$ is isomorphic to
$(\mathbf{C}^{N+1}\backslash \{(0,0,\ldots,0)\})/\sim$ as sets,
where the equivalence relation $\sim$ is defined by
$
(x_{0},\ldots,x_{N})\sim(y_{0},\ldots,y_{N})\Leftrightarrow\exists
\lambda\in \mathbf{C}-0$ such that
$\lambda x_{i}=y_{i} \forall
~0\leq i\leq N$.
Let $R=\mathbf{C}[\alpha_{0},\alpha_{1},\ldots, \alpha_{N}]$ be the
ring of polynomial in $N+1$ variables over complex number field
$\mathbf{C}$ , $R_{d}$ be the set of all homogeneous polynomials of
degree $d$, and $I_{d}$ be the set of polynomials in $R_{d}$. For
an ideal $I$ of $R$, we have also $I_{d}=I\cap R_{d}$.
 Then,
$I=\bigoplus^{\infty}_{d=0}I_{d}$ is called the homogeneous ideal of
$R$. If a prime ideal $P$ is homogeneous, then $P$ is called
 homogeneous prime ideal.
Now, we define
\begin{equation}
\mathbf{P}^{N}_{\mathbf{C}}=\{P:~P~ is ~ a ~ homogeneous~ prime~
ideal ~ of~R~and~P\neq(\alpha_{0},\ldots,
\alpha_{N})\}
\end{equation}
and by defining $V(I)=\{P\in\mathbf{P}^{N}_{\mathbf{C}}:I\in P\}$ as
a closed set of $\mathbf{P}^{N}_{\mathbf{C}}$ we get a topology
on $\mathbf{P}^{N}_{\mathbf{C}}$.
Now, we will illustrate this construction by explicitly discuss the
space of a quantum bit (qubit) $ \ket{\Psi}=\alpha_{0}
\ket{0}+\alpha_{1}\ket{1}$. Let
$$U_{0}=
(Spec\mathbf{C}[\alpha_{0}],\mathcal{O}_{Spec\mathbf{C}[\alpha_{0}]})~
and~U_{1}=(Spec\mathbf{C}[\alpha_{1}],
\mathcal{O}_{Spec\mathbf{C}[\alpha_{1}]}).$$
  Then we can define an affine scheme structure on an open set
$X_{\alpha_{0}}=D(\alpha_{0})$ of
$X=U_{0}=\mathrm{Spec}\mathbf{C}[\alpha_{0}]$ as
\begin{equation}
U_{01}=(\mathrm{Spec}\mathbf{C}[\alpha_{0},1/\alpha_{0}],
\mathcal{O}_{\mathrm{Spec}\mathbf{C}
[\alpha_{0},1/\alpha_{0}]}),
\end{equation}
 where $\mathcal{O}_{\mathrm{Spec}\mathbf{C}[\alpha_{0},1/\alpha_{0}]}
=\mathcal{O}_{X}|X_{\alpha_{0}}$
Moreover, we define
$D(\alpha_{1})$ in similar manner.
Furthermore, the isomorphic
$\phi:
\mathbf{C}[\alpha_{1},1/\alpha_{1}]\longrightarrow\mathbf{C}[\alpha_{0},1/\alpha_{0}]$
defined by $f(\alpha_{1},1/\alpha_{1})\longmapsto
f(\alpha_{0},1/\alpha_{0})$ induces an isomorphism of affine scheme
$(\phi^{c},\phi^{\diamond}):U_{01}\longrightarrow U_{10}=(Spec\mathbf{C}[\alpha_{1},1/\alpha_{1}],
\mathcal{O}_{Spec\mathbf{C}
[\alpha_{1},1/\alpha_{1}]}).$
  Then by
gluing $U_{0}$ and $U_{1}$ through this isomorphism gives the scheme
$\mathbf{P}^{1}_{\mathbf{C}}=(Z,\mathcal{O}_{Z})$, where
$Z=X\bigcup_{\phi^{c}}Y$ is obtained by gluing $X$ and $Y$ by
identifying the open sets $X_{\alpha_{0}}$ and $Y_{\alpha_{1}}$ by
the map $\phi^{c}$.
 We have also $\mathcal{O}_{Z}|X=\mathcal{O}_{X}$
and $\mathcal{O}_{Z}|Y=\mathcal{O}_{Y}$. Thus the structure sheaf
$\mathcal{O}_{Z}$ is obtained by identifying
$\mathcal{O}_{X}|X_{\alpha_{0}}$ and
$\mathcal{O}_{Y}|Y_{\alpha_{1}}$ by $\phi^{\diamond}$.

We can also generalize the above construction as follows. A
  commutative ring
$S=\bigoplus^{\infty}_{d=0}S_{d}$ that satisfies $S_{d}S_{s}\subset
S_{d+s}$ is called graded ring.
For example let
$S=\mathbf{C}[\alpha_{0},\alpha_{1},\ldots, \alpha_{N}]$ be the ring
of polynomial in $N+1$ variables over complex number field
$\mathbf{C}$.
 Moreover, let
 \begin{equation}
 S_{d}= \{F\in S:~ F ~be ~a ~homogeneous ~polynomials ~of
~degree ~d\}.
 \end{equation}
 Then $S=\bigoplus^{\infty}_{d=0}S_{d}$ is a graded ring and an
ideal $I=\bigoplus^{\infty}_{d=0}I_{d}$ of $S$, where $I_{d}=I\cap
S_{d}$ is called a homogeneous ideal.

Note that, an ideal $I$ is
homogeneous if and only if for an arbitrary element
$F=\bigoplus^{\infty}_{d=1}F_{d}$, we have $F_{d_{i}}\in I$ for all
$i=1,2,\ldots, l$.
 If $I$ is prime, then is called the homogeneous
prime ideal of $S$.
 For example, $S_{+}=\bigoplus^{\infty}_{d\geq
1}S_{d}$ is a homogeneous ideal of the graded ring $S$. Next, we
define
\begin{equation}
\mathrm{Proj}S=\{P:~P~is ~a ~homogeneous ~prime ~ideal~
of~S~and~P ~is ~ not ~ in ~ S_{+}\},
\end{equation}
to be the homogeneous prime spectrum of $S$.
Now, the Zariski
topology can be defined on $\mathrm{Proj}S$ by taking
$V(Q)=\{P\in\mathrm{Proj}S: P\supset Q\}$ as closed set.

Next, we will investigate the geometrical structure of
multi-qubit quantum states
$\ket{\Psi}=\sum^{1}_{i_{1}=0}\sum^{1}_{i_{2}=0}\cdots\sum^{1}_{i_{m}=0}
\alpha_{i_{1}i_{2}\cdots i_{m}} \ket{i_{1}i_{2}\cdots i_{m}}$ based
complex projective scheme. Our construction is based on a map called
the Segre morphism. Let us consider the following map
\begin{equation}\nonumber
\begin{array}{ccc}\label{segmap}
\mathbf{P}^{1}_{\mathbf{C}}\times\mathbf{P}^{1}_{\mathbf{C}}
\times\cdots\times\mathbf{P}^{1}_{\mathbf{C}}&\longrightarrow&
\mathbf{P}^{2^{m}-1}_{\mathbf{C}}=
\mathrm{Proj}\mathbf{C}[\alpha_{00\cdots0},\alpha_{0\cdots01},\ldots,\alpha_{1\cdots11}]\\
\end{array}
\end{equation}
\begin{equation}\nonumber
\begin{array}{ccc}\label{segmap}
 ((\alpha^{0}_{0}:\alpha^{0}_{1}),\ldots,(\alpha^{m-1}_{0}:\alpha^{m-1}_{1})) & \longmapsto&
 (\ldots,\alpha^{0}_{i_{1}}\alpha^{1}_{i_{2}}\cdots\alpha^{m-1}_{i_{m}},\ldots). \\
\end{array}
\end{equation}
where  $\alpha_{i_{1}i_{2}\cdots i_{m}}$,$0\leq i_{j}\leq 1$ are
homogeneous coordinate-functions on
$\mathbf{P}^{2^{m}-1}_{\mathbf{C}}$.
Moreover, let
$\mathbf{X}_{1}=\mathrm{Proj}\mathbf{C}[\beta^{1}_{0}:\beta^{1}_{1}]=
\mathbf{P}^{1}_{\mathbf{C}},
\ldots,
\mathbf{X}_{m}=
\mathrm{Proj}\mathbf{C}[\beta^{m-1}_{0}:\beta^{m-1}_{1}]=\mathbf{P}^{1}_{\mathbf{C}}$
and  $f_{j}:\mathbf{X}_{j}\longrightarrow\mathrm{Spec}\mathbf{C}$
for all $j=1,2,\ldots,m$.
 Furthermore, let
$$g:\mathbf{Z}=\mathbf{X}_{1}\times_{\mathrm{Spec}\mathbf{C}}
\mathbf{X}_{2}\times_{Spec
\mathbf{C}}\cdots\times_{Spec\mathbf{C}}\mathbf{X}_{m}
\longrightarrow Spec\mathbf{C}$$ be the structure
morphism and\begin{equation}
p_{j}:\mathbf{Z} \longrightarrow\mathbf{X}_{j}
\end{equation}
 be the projection for
$j=1,2,\ldots,m$.
So, we have $\Gamma(\mathbf{X}_{j},
\mathcal{O}_{\mathbf{X}_{j}}(1))=\mathbf{C}\beta^{j}_{0}\oplus\mathbf{C}\beta^{j}_{0}=V_{j}$.
Now, the natural $\mathcal{O}_{\mathbf{X}}$-homomorphism
\begin{equation}
\gamma_{j}:f^{*}_{j}V_{j}=\mathcal{O}_{\mathbf{X}_{j},\beta^{j}_{0}}
\oplus\mathcal{O}_{\mathbf{X}_{j},\beta^{j}_{0}}\longrightarrow\mathcal{O}_{\mathbf{X}_{j}}(1)
\end{equation}
is surjective and
$\phi_{(\mathcal{O}_{\mathbf{X}_{j}}(1),\gamma_{j})}:\mathbf{X}_{j}
\longrightarrow\mathbf{P}(V_{j})$
is an isomorphism over $\mathbf{C}$.
  Next we construct the
invertible sheaf $\mathcal{L}$ over $Z$ by
\begin{equation}
 \mathcal{L}=p^{*}_{1}\mathcal{O}_{\mathbf{X}_{1}}(1)\otimes
p^{*}_{2}\mathcal{O}_{\mathbf{X}_{2}}(1)\otimes\cdots\otimes
p^{*}_{m}\mathcal{O}_{\mathbf{X}_{m}}(1).
\end{equation}
 Then, we have also
the following surjective homomorphism
\begin{equation}
    \gamma:g^{*}(V_{1}\otimes
    V_{2}\otimes\cdots\otimes V_{m})=p^{*}_{1}f^{*}_{1}(V_{1})\otimes
    p^{*}_{2}f^{*}_{2}(V_{2})\otimes\cdots\otimes
    p^{*}_{m}f^{*}_{m}(V_{m})\rightarrow\mathcal{L}.
\end{equation}
This construction gives us a scheme morphism\begin{equation}
\phi_{(\mathcal{L},\gamma)}:Z\longrightarrow \mathbf{P}(V_{1}\otimes
V_{2}\otimes\cdots\otimes
V_{m})\simeq\mathbf{P}^{2^{m}-1}_{\mathbf{C}},
\end{equation}
since $V_{1}\otimes
V_{2}\otimes\cdots\otimes V_{m}$ is isomorphic to
$\mathbf{C}^{2^{m}}$.
Then one can show that the map
$Z(\mathbf{C})=\mathbf{P}(V_{1}\otimes V_{2}\otimes\cdots\otimes
V_{m})(\mathbf{C})=\mathbf{P}^{2^{m}}(\mathbf{C})$. Let us consider the map $\psi_{j}:V_{j}\longrightarrow \mathbf{C}$.
Then the $\mathbf{C}$-value point of the space $\mathbf{X}_{j}$ are
in one-to-one correspondence with the equivalence classes of
non-zero map $\psi_{j}$ which is uniquely determined by
$\psi_{j}(\beta^{j-1}_{0})=\alpha^{j-1}_{0}$.

Thus $
((\alpha^{0}_{0}:\alpha^{0}_{1}),\ldots,(\alpha^{m-1}_{0}:\alpha^{m-1}_{1}))$
corresponds to an element of
$\mathbf{X}_{1}(\mathbf{C})\times
\mathbf{X}_{2}(\mathbf{C})\times\cdots \times
\mathbf{X}_{m}(\mathbf{C})$ in one-to-one way.
    That is, the image
of $
((\alpha^{0}_{0}:\alpha^{0}_{1}),\ldots,(\alpha^{m-1}_{0}:\alpha^{m-1}_{1}))$
under the morphism
\begin{equation}
\phi_{(\mathcal{L},\gamma)}:Z\longrightarrow \mathbf{P}(V_{1}\otimes
V_{2}\times\cdots\otimes V_{m})
\end{equation}
 corresponds
to the equivalent class of
\begin{equation}
\psi=\psi_{1}\otimes\psi_{2}\times\cdots\otimes\psi_{m}:V_{1}\otimes
V_{2}\times\cdots\otimes V_{m}\longrightarrow \mathbf{C}.
\end{equation}
This map is uniquely determined by
\begin{equation}
\begin{array}{ccc}
\psi(\beta^{0}_{0}\otimes
\beta^{1}_{0}\otimes\cdots\otimes\beta^{m-1}_{0})&=&\alpha^{0}_{0}\alpha^{1}_{0}\cdots
\alpha^{m-1}_{0}\equiv\alpha_{00\cdots0}\\
  \psi(\beta^{0}_{0}\otimes
\beta^{1}_{0}\otimes\cdots\otimes\beta^{m-1}_{1})&=&\alpha^{0}_{0}\alpha^{1}_{0}\cdots
\alpha^{m-1}_{1} \equiv\alpha_{0\cdots01}\\
 &\vdots&\\
\psi(\beta^{0}_{1}\otimes
\beta^{1}_{1}\otimes\cdots\otimes\beta^{m-1}_{1})&=&\alpha^{0}_{1}\alpha^{1}_{1}\cdots
\alpha^{m-1}_{1}\equiv\alpha_{11\cdots1}
\end{array}
\end{equation}
     which is the equivalent class of $\psi$ corresponding to
\begin{equation}
(\alpha^{0}_{0}\alpha^{1}_{0}\cdots
\alpha^{m-1}_{0},\alpha^{0}_{0}\alpha^{1}_{0}\cdots
\alpha^{m-1}_{1},\ldots,\alpha^{0}_{1}\alpha^{1}_{1}\cdots
\alpha^{m-1}_{1})
=
(\alpha_{00\cdots0},\alpha_{0\cdots01},\ldots,\alpha_{1\cdots11}).
\end{equation}
Thus the map $\phi_{(\mathcal{L},\gamma)}$ is exactly the Segre map
(\ref{segmap}) and the space of separable state of a multi-qubit
quantum state is give by  $\mathrm{Im}\phi_{(\mathcal{L},\gamma)}$
the image of $\phi_{(\mathcal{L},\gamma)}$
\begin{equation}
\mathrm{Im}\phi_{(\mathcal{L},\gamma)}=\{\alpha_{i_{1}i_{2}\cdots
i_{m}}\in\mathbf{P}^{2^{m}-1}_{\mathbf{C}}:
~\alpha_{[k_{1}k_{2}\cdots k_{m}}\alpha_{l_{1}l_{2}\cdots
l_{m}]}=0,~0\leq i_{j}\leq 1,~ i=,k,l\}.
\end{equation}
  The measure of entanglement for multi-qubit states can also be constructed based on quadratic polynomial defining  the Segre
morphism which also coincides with well-known concurrence for
two-qubit and three-qubit states \cite{Hosh1,Hosh2}.
     There is also another
method that generalize the above map based on quasicoherent sheaf
which we will not  discuss here due to lack of space.


\subsection{Example: a two-qubit state }
Now, we illustrate this construction by working out in detail a
non-trivial example of quantum system, namely a two qubit state. The
image of the map
$\mathbf{P}^{1}_{\mathbf{C}}\times\mathbf{P}^{1}_{\mathbf{C}}\longrightarrow\mathbf{P}^{3}_{\mathbf{C}}
=\mathrm{Proj}\mathbf{C}[\alpha_{00},\alpha_{01},\alpha_{10},\alpha_{11}]$
defined by
\begin{equation}
((\alpha^{0}_{0}:\alpha^{0}_{1}),(\alpha^{1}_{0}:\alpha^{1}_{1}))
\longmapsto(\alpha^{0}_{0}\alpha^{1}_{0},\alpha^{0}_{0}\alpha^{1}_{1},\alpha^{0}_{1}\alpha^{1}_{0},
\alpha^{0}_{1}\alpha^{1}_{1})
\end{equation}
 is a quadratic surface in
$\mathbf{P}^{3}_{\mathbf{C}}$.  Now, we want to construct this
surface in terms of schemes. Let,
$\mathbf{X}_{1}=\mathrm{Proj}\mathbf{C}[\beta^{0}_{0}:\beta^{0}_{1}]=\mathbf{P}^{1}_{\mathbf{C}}$
and
$\mathbf{X}_{2}=\mathrm{Proj}\mathbf{C}[\beta^{1}_{0}:\beta^{1}_{1}]=\mathbf{P}^{1}_{\mathbf{C}}$.
Moreover, let
$f_{1}:\mathbf{X}_{1}\longrightarrow\mathrm{Spec}\mathbf{C}$ and
$f_{1}:\mathbf{X}_{1}\longrightarrow\mathrm{Spec}\mathbf{C}$.
Furthermore, let
$g:\mathbf{Z}=\mathbf{X}_{1}\times_{\mathrm{Spec}\mathbf{C}}\mathbf{X}_{2}\longrightarrow\mathrm{Spec}\mathbf{C}$
be the structure morphism and
$p_{j}:\mathbf{Z}=\mathbf{X}_{1}\times_{\mathrm{Spec}\mathbf{C}}\mathbf{X}_{2}
\longrightarrow\mathbf{X}_{j}$ be the projection for $j=1,2$.
So, we have $\Gamma(\mathbf{X}_{1},
\mathcal{O}_{\mathbf{X}_{1}}(1))=\mathbf{C}\alpha_{0}\oplus\mathbf{C}\alpha_{1}=V_{1}$
and $\Gamma(\mathbf{X}_{2},
\mathcal{O}_{\mathbf{X}_{2}}(1))=\mathbf{C}\beta_{0}\oplus\mathbf{C}\beta_{1}=V_{2}$.
Now, the natural $\mathcal{O}_{\mathbf{X}}$-homomorphism
\begin{equation}
\gamma_{1}:f^{*}_{1}V_{1}=\mathcal{O}_{\mathbf{X}_{1},\alpha^{0}_{0}}
\oplus\mathcal{O}_{\mathbf{X}_{1},\alpha^{0}_{1}}\longrightarrow\mathcal{O}_{\mathbf{X}_{1}}(1)
\end{equation}
is surjective and
$\phi_{(\mathcal{O}_{\mathbf{X}_{1}}(1),\gamma_{1})}:\mathbf{X}_{1}\longrightarrow\mathbf{P}(V_{1})$
is an isomorphism over $\mathbf{C}$. Next we construct the
invertible sheaf over $Z$ by
$\mathcal{L}=p^{*}_{1}\mathcal{O}_{\mathbf{X}_{1}}(1)\otimes
p^{*}_{2}\mathcal{O}_{\mathbf{X}_{2}}(1)$. Then we have also
following surjective homomorphism
\begin{equation}
    \gamma:g^{*}(V_{1}\otimes
    V_{2})=p^{*}_{1}f^{*}_{1}(V_{1})\otimes
    p^{*}_{2}f^{*}_{2}(V_{2})\longrightarrow\mathcal{L}.
\end{equation}
This construction gives us a scheme morphism
$\phi_{(\mathcal{L},\gamma)}:Z\longrightarrow
\mathbf{P}(V_{1}\otimes V_{2})\simeq\mathbf{P}^{3}_{\mathbf{C}}$,
since $V_{1}\otimes V_{2}$ is isomorphic to $\mathbf{C}^{4}$. Then
one can show that the map $Z(\mathbf{C})=\mathbf{P}(V_{1}\otimes
V_{2})=\mathbf{P}^{3}(\mathbf{C})$, see also ref. \cite{Ueno}.
Let us consider the map $\psi_{1}:V_{1}\longrightarrow \mathbf{C}$.
Then the $\mathbf{C}$-value point of the space $\mathbf{X}_{1}$ are
in one-to-one correspondence with the equivalence classes of
non-zero map $\psi_{1}$ which is uniquely determined by
$\psi_{1}(\beta^{0}_{0})=\alpha^{0}_{0}$ and
$\psi_{1}(\beta^{0}_{1})=\alpha^{0}_{1}$. We can also get a similar
construction for the $\mathbf{C}$-value point of the space
$\mathbf{X}_{2}$. Thus
$((\alpha^{0}_{0},\alpha^{0}_{1}),(\alpha^{1}_{0},\alpha^{1}_{1}))$
corresponds to an element of $\mathbf{X}_{1}(\mathbf{C})\times
\mathbf{X}_{2}(\mathbf{C})$ in one-to-one way. That is the image of
$((\alpha^{0}_{0},\alpha^{0}_{1}),(\alpha^{1}_{0},\alpha^{1}_{1}))$
under the morphism $\phi_{(\mathcal{L},\gamma)}:Z\longrightarrow
\mathbf{P}(V_{1}\otimes V_{2})$ corresponds to the equivalent class
of $\psi=\psi_{1}\otimes\psi_{2}:V_{1}\otimes V_{2}\longrightarrow
\mathbf{C}$. This map is uniquely determined by
$$
\begin{array}{cc}
  \psi(\beta^{0}_{0}\otimes
\beta^{0}_{1})=\alpha^{0}_{0}\alpha^{1}_{0}\equiv\alpha_{00} &~~~~
\psi(\beta^{0}_{0}\otimes
\beta^{1}_{1})=\alpha^{0}_{0}\alpha^{1}_{1}\equiv\alpha_{01} \\
  \psi(\beta^{0}_{1}\otimes
\beta^{1}_{0})=\alpha^{0}_{1}\alpha^{1}_{0}\equiv\alpha_{10} &~~~~
\psi(\beta^{0}_{1}\otimes
\beta^{1}_{1})=\alpha^{0}_{1}\alpha^{1}_{1}\equiv\alpha_{11}
\end{array}
$$
 which is the equivalent
class of $\psi$ corresponding to
$(\alpha^{0}_{0}\alpha^{1}_{0},\alpha^{0}_{0}\alpha^{1}_{1},
\alpha^{0}_{1}\alpha^{1}_{0},\alpha^{0}_{1}\alpha^{1}_{1})=
(\alpha_{00},\alpha_{01},\alpha_{10},\alpha_{11})$. Thus the map
$\phi_{(\mathcal{L},\gamma)}$ is exactly the Segre map and the image
of $\phi_{(\mathcal{L},\gamma)}$ is the space of separable
 two-qubit quantum
states
$\Psi=\alpha_{00}\ket{00}+\alpha_{00}\ket{01}+\alpha_{10}\ket{10}+\alpha_{11}\ket{11}$,
that is
\begin{equation}
\mathrm{Im}\phi_{(\mathcal{L},\gamma)}=\{\alpha_{k_{1}k_{2}}\in\mathbf{P}^{3}_{\mathbf{C}}
: ~\alpha_{00}\alpha_{11}=\alpha_{01}\alpha_{10}, ~~k_{1},k_{2}=1,2\}
\end{equation}
The measure of entanglement for such state can also be constructed
by using the definition of the Segre morphism which also coincides
with well-known concurrence.


\begin{flushleft}
\textbf{Acknowledgments:}  The  work was supported  by the Swedish Research Council (VR).
\end{flushleft}

\end{document}